\documentclass[prl,showpacs,showkeys,twocolumn]{revtex4}
\usepackage{amsmath}
\usepackage{amssymb}
\usepackage{float}
\usepackage{bm,times}
\usepackage{graphicx}

\newcommand{\vect}[1] {\mathbf{#1}}

\newcommand{\up} {\uparrow}
\newcommand{\down} {\downarrow}

\newcommand{\brac}[1] {\langle #1\rvert}
\newcommand{\ket}[1] {\lvert #1\rangle}

\begin{document}

\title{A new many-body wave function for BCS-BEC crossover in Fermi gases}
\author{Shina~Tan}
\author{K.~Levin}
\affiliation{James Franck Institute and Department of Physics,
  University of Chicago, Chicago, Illinois 60637}

\begin{abstract}
We present a new many body formalism for BCS-BEC crossover, which
represents a modification of the BCS-Leggett
ground state to
include 4-fermion, and higher correlations.
In the BEC regime, we show how our approach
contains 
the \textit{Petrov et al} 4-fermion behavior and
associated scattering length
$a_{dd}$ at short distances,
and secondly reduces to composite-boson Bogoliubov physics
at long distances.
It reproduces the Lee-Yang term, whose
numerical value is also fixed by $a_{dd}$.
We have also examined the next term beyond the Lee-Yang correction
in a phenomenological fashion, building on cloud size data
and collective mode experiments.  However, one has to view this phenomenological
analysis
with some caution since experiments are in a state of
flux and are performed close to
unitarity.
\end{abstract}


\maketitle

There have been exciting new developments in the field
of trapped fermionic atoms
\cite{background, collectivemodes, Zwierlein_vortices}
in which 
applied magnetic fields
are used to effect a transition from BCS
superfluidity to Bose Einstein condensation (BEC).  
The simplest mean field wave function \cite{Eagles1969PR,Leggett1980article}
which describes these
phenomena at $T=0$ is, effectively, the BCS wave function 
$\ket{\psi_\text{BCS}}$
with self consistently determined (fermionic) chemical
potential in the presence of pairing interactions of arbitrary
strength.
This has met with some experimental
success \cite{ChenRev} but it is clearly lacking
in the sense that it does not incorporate the exact two-molecule (4-fermion) scattering physics \cite{Petrov2004PRL}.
In this mean field theory, the 4-fermion
interaction, parametrized by the dimer-dimer scattering
length $a_{dd}$, is overestimated by a factor of 3, leading to
a similar overestimate in the cloud size
\cite{Bartenstein2004cloudsize}.
Condensate depletion is also absent in this lowest order mean-field theory.

This earlier work raises a fundamental
question which we address here.
How does one arrive at a many body wave function
which incorporates the
few body constraints and
what are its
implications for experiment. 
While, there have been
efforts \cite{Holland,Strinati3} to go beyond the leading order mean
field theory in the literature, there has never
been a demonstrated consistency between the exact few body constraints and
N-body behavior.
Our work is the first to make these
two points of view compatible and in the process, introduces
a new perturbation scheme which replaces the
usual Feynmann diagrammatic expansion.
Here we derive an equation of state for molecular bosons
at the level of the Bogoliubov approximation, and, thereby 
set up a basis from which one can compute various properties of the
gas. Of particular interest are
those which are associated with the underlying fermionic physics,
which are absent
in a conventional BEC of true bosons.

The parameters appearing in our wave function are determined directly by an energy minimization
procedure; we stress that there are no fitting parameters.
The equation of state reduces to the well known
Lee-Yang like formula \cite{LeeYang1957}, but with the scattering length $a_{dd}$ determined
by the exact 4-fermion physics \cite{Petrov2004PRL}.
Finally, 
our wave function appears compatible with the general features of
experimental data on two different experiments: cloud sizes 
\cite{Bartenstein2004cloudsize}
and collective modes 
\cite{Kinast2004PRA}.

%

We begin by summarizing our results for the many body wave function
and its consequences before providing the detailed methodology
we employ. Additional details are given in
the unpublished supporting material (Ref.~\cite{_Tan}) which we have
made available.
One may rewrite the mean field wave function
\cite{Eagles1969PR, Leggett1980article}
as 
\begin{equation}\label{eq:BCS2}
\ket{\psi_\text{BCS}}\propto \exp\Bigl(\sum_\vect k\alpha_{\vect k}^{} c_{\vect k\up}^\dagger
c_{-\vect k\down}^\dagger\Bigr)\ket{0},
\end{equation}
where $c_{\vect k\sigma}$ destroys fermions
with momentum $\hbar\vect k$ and spin $\sigma$ \cite{_spin}, and
$\ket{0}$ is the particle vacuum.
Written in this way, the BCS wave function is seen to be the ``composite-boson" version of the Gross-Pitaevskii (GP)
wave function of conventional BEC. 

It seems natural to presume that corrections to this GP-like wave function
should capture some of the physics of the conventional
Bogoliubov approach which provides an improvement over 
the GP wave function, appropriate to bosonic systems.  
We may rewrite the
corresponding Bogoliubov wave function as
\begin{equation}
\ket{\psi_\text{Bogoliubov}}=\exp\biggl(N_0^{1/2}b_0^\dagger+\sum_{\vect q>0}
x_\vect q b_{\vect q}^\dagger b_{-\vect q}^\dagger\biggr)\ket{0},
\label{eq:bogol}
\end{equation}
where $b_\vect q^\dagger$ is the creation operator of bosons with momentum $\hbar \vect q$
which, like $b_0^\dagger$, must be replaced by a bilinear form of fermion creation operators.
In this way, one sees that four fermion terms are required
to go beyond the simple BCS-like mean-field theory.

On this basis we return to Eq.~\eqref{eq:BCS2}
and write the natural correction terms.  As will be shown
below, to capture the physics of the deep BEC region, 6-fermion
as well as 4-fermion corrections are needed. This
goes beyond earlier work \cite{Holland}.
More
generally one can contemplate a many body wave function of 
the form
\begin{widetext}
\begin{equation}\label{eq:psi}
\ket{\psi}=\exp\biggl(\frac{1}{2!}
\sum_{\vect K}\alpha_{\vect K}^{}c_{\vect K}^\dagger c_{-\vect K}^\dagger
+\frac{1}{4!}\sum_{\vect K\text{'s}}
\beta_{\vect K_1 \vect K_2 \vect K_3 \vect K_4}^{}
c_{\vect K_1}^\dagger c_{\vect K_2}^\dagger c_{\vect K_3}^\dagger c_{\vect K_4}^\dagger
+\frac{1}{6!}\sum_{\vect K\text{'s}}
\gamma_{\vect K_1\cdots\vect K_6}^{}
c_{\vect K_1}^\dagger\dots c_{\vect K_6}^\dagger+\cdots
\biggr)\ket{0},
\end{equation}\end{widetext}
where each $\vect K_i$ represents
a shorthand notation for $\vect k_i\sigma_i$, and $-\vect K$ refers
to the fact
that both the momentum and the spin are
reversed. 
There are constraints imposed on the coefficients
$\alpha$, $\beta$, $\gamma$, etc, 
which can be readily formulated \cite{_Tan}, although we will
not do so here.

We show below
how to compute physical properties in terms of
these wave function parameters. We do this in a fashion which is also consistent
with 
Ref.~\cite{Petrov2004PRL}.
Thus, to determine, for example, $a_{dd}$, the
wave function is adjusted
to minimize the energy, at a given fixed number of fermions. For general
values of $k_Fa$ this procedure is very difficult to implement.
However,
in the \textit{low-density} regime, given more
precisely by $n_d a_{dd}^3\ll1$,
higher-order correlations in the wave function are much
weaker, and we are able to construct a \textit{systematic}
perturbation theory. For this reason, our theory is mostly limited
to the deep BEC regime of the crossover ($n_d a_{dd}^3\ll1$).          This is different from
some other approaches which produce various smooth curves of the ground state energy
in the whole BCS-BEC crossover \cite{ChenRev, Strinati2005PRB3page, Hu2006}.
Despite this limitation, physical observables (such as the equation
of state) take the form of well-controlled low-density expansions,
and the functional forms, as well as the coefficients of such expansions
can be exactly determined within our theory, whereas the alternative
theories \cite{ChenRev, Strinati2005PRB3page, Hu2006} lack such exactness in the BEC limit.

The ground state expectation value of any physical observable $O$ is
\begin{equation}\label{eq:expectation}
\langle O\rangle
=\frac{\brac{0}\mathcal{T}O
\exp(\alpha^\dagger+\beta^\dagger+\cdots+\alpha+\beta+\cdots)\ket{0}}
{\brac{0}\mathcal{T}
\exp(\alpha^\dagger+\beta^\dagger+\cdots+\alpha+\beta+\cdots)\ket{0}},
\end{equation}
where $\alpha$, $\beta$, etc, are the terms in the exponent of the right side
of Eq.~\eqref{eq:psi}, and 
$\mathcal{T}$ is a ``time-ordering" operator defined
by assigning the fermion creation
operators in $\alpha$, $\beta$, etc the earliest ``time",
the annihilation operators in $\alpha^\dagger$, $\beta^\dagger$, etc
the latest ``time". Finally, those times appearing in
$O$ enter as intermediate
``times" increasing monotonically from right to left.
A standard Feynman diagrammatic method
is then used to evaluate Eq.~\eqref{eq:expectation}. 
It should be noted that, in our present approach,
the terms in the \textit{wave function} (instead of the
interaction) serve as
vertices, in contrast to the textbook applications of Feynman diagrams.
Our approach is suited to the bosonic regime,
where the attractive interaction between fermions is so \emph{strong}
that they form bound states.

One can contemplate very general interactions between two fermions
which lead to these bound states.
Under these circumstances, 
it follows from Eq.~\eqref{eq:psi}
that the ground state energy is given by 
\begin{multline}\label{eq:E}
E/\Omega=n_dE_d^{(0)}+2\pi\hbar^2 n_d^2 a_{dd}/m_d \\\times\Bigl\{
1+[128/(15\sqrt{\pi})]\sqrt{n_d a_{dd}^3}+\cdots\Bigr\},
\end{multline}
which is similar to that of point-like bosons \cite{LeeYang1957}.
Here $\Omega$ is the volume, $n_d=n/2$ is the number density of dimers,
$m_d=2m$ is the mass of the dimer, $m$ is the mass of each fermion,
$E_d^{(0)}$ the energy of a zero-momentum
dimer in vacuum, and $a_{dd}$ is the self consistently
determined dimer-dimer scattering length.
It should be stressed that this equation is very general; it applies 
even when the fermionic interaction is
\textit{not} a contact interaction.
When the interaction has a finite range (comparable to the fermionic scattering length $a$), the famous calculation of Petrov et al in Ref.~\cite{Petrov2004PRL}
needs to be corrected and $a_{dd}/a\neq0.6$.
It is the \textit{actual} 
$a_{dd}$, determined from the solution to the quantum 4-fermion problem
for the low-energy scattering of two molecules [see Eq.~\eqref{eq:phi4} below], that must be used in Eq.~\eqref{eq:E}.
When the interaction is of very short range ($\ll a$) then the $a_{dd}$
in our theory reduces to that in Ref. \cite{Petrov2004PRL}.

An equation of state associated with the
first two terms in Eq.~\eqref{eq:E} (between the curly brackets)
 was first proposed in Ref. \cite{Stringari2004Eu},
based on the assumption that fermionic dimers
were governed by the same equation of state as point-like bosons. 
It is of considerable interest that we have confirmed
this equation of state in the context of a fermionic
system.
Also note that this equation of state
[Eq.~\eqref{eq:E}] contains 
$n_d a_{dd}^3\ll1$ as a small parameter \cite{_a_dd}. 
While one might be inclined to consider a
power series in $k_{F}a$, in this low density regime,
where $k_{F}\propto n^{1/3}$ 
is the formal Fermi wave vector, we have shown that
this is incorrect.
When all fermions form tightly bound pairs, there is no
Fermi sphere, and no natural
expansion in terms of $n^{1/3}$ exists.


We now provide some details for our
calculation by sketching
how this wave function is determined in the BEC regime.
If $n_da_{dd}^3\ll1$, higher correlation terms in the exponent in Eq.~\eqref{eq:psi}
are progressively less important. Their orders of magnitude are derived,
determining which coefficients
we should retain, at any given order in the expansion of the equation of state.
\emph{At the leading order}, only $\alpha_\vect K$ is important, and 
the two-body Schrodinger
equation is reproduced upon minimization of $E-\mu N$ ($N$ is the number of fermions). This wave function
corresponds to 
a collection of dimers.
\emph{At the next order}, $\beta$ enters and is adjusted to minimize energy. It is consequently found that $\beta$
satisfies the 4-body Schrodinger equation for the low-energy scattering of two molecules at the leading order.
We thus use the solution to the latter equation to construct $\beta$ approximately [the result is shown in Eq.~\eqref{eq:beta1234}
below]. In the process $\alpha_\vect K$ is computed to a higher order, to minimize energy at this higher level of approximation.
Summing up all the contributions to energy up to this level (using the results for $\alpha$ and $\beta$ that one just obtained),
 one deduces a mean-field interaction between dimers;
note that here one derives the correct $a_{dd}$ in accordance with Ref.~\cite{Petrov2004PRL} in such interaction term.

\emph{At the third order},
$\gamma_{\vect K_1+\vect q/4,\cdots,\vect K_4+\vect q/4,\vect k'-\vect q/2\up,-\vect k'-\vect q/2\down}$
now appears. This order corresponds to the
usual Lee-Yang term \cite{LeeYang1957} for
point bosons, although here the derivation is for
the molecular boson case. Similarly,
$\alpha_\vect K$ and $\beta$ are 
recomputed to yet a higher order of accuracy. 
Corrections to the latter are now predominantly at small $q$ $\bigl[\sim O(\sqrt{n_da_{dd}})\bigr]$.

At this third order,
the function $\beta$ in Eq.~\eqref{eq:psi} \textit{unifies}
the Bogoliubov theory of composite bosons and the exact
4-fermion calculations of  Ref.~\cite{Petrov2004PRL}.
We show this as follows by deducing an asymptotic formula
from Bogoliubov theory for composite bosons. We
introduce 
$b_{\vect q}^\dagger =\sum_{\vect k}
\phi_{\vect k} c_{\vect q/2+\vect k\up}^\dagger c_{\vect q/2-\vect k\down}^\dagger$, where
$\phi_{\vect k}$ is the internal wave function of the dimer,
satisfying
$\sum_{\vect k}\lvert\phi_\vect k\rvert
^2=1$.
With this substitution, it follows that
\begin{equation}\label{eq:beta_smallq}
\beta_{\vect q/2+\vect k\up, \vect q/2-\vect k\down, -\vect q/2+\vect k'\up,
-\vect q/2-\vect k'\down} \approx
x_\vect q\phi_\vect k\phi_{\vect k'}.
\end{equation}
In fact, by strictly applying the energy minimization principle alone, 
without resorting to the picture of bosons, we are able to derive the same formula as above
(under the condition $q\ll 1/r_d$ and $q\ll\lvert\vect k-\vect k'\rvert$, where $r_d\sim a$ is the molecular size), and our formula
for $x_\vect q$ coincides, at the leading order, with that of a traditional Bose gas \cite{LeeYang1957} whose scattering length equals $a_{dd}$.
For
$1/r_d\gg q\gg\sqrt{n_da_{dd}}$ and $\lvert\vect k-\vect k'\rvert \gg q$, our theory yields the standard asymptotic
formula for 
$x_\vect q$
to obtain
\begin{equation}
\beta_{\vect q/2+\vect k\up,\vect q/2-\vect k\down,-\vect q/2+\vect k'\up,-\vect q/2-\vect k'\down}
=-\frac{4\pi n_da_{dd}}{q^2}\phi_\vect k\phi_{\vect k'}
\label{eq:6}
\end{equation}
plus higher order corrections.

Now let us compute the related behavior which derives from
the nonperturbative four-body physics of Ref.~\cite{Petrov2004PRL} .
When $q$ and $\lvert \vect k-\vect k'\rvert$ are both $\gg\sqrt{n_da_{dd}}$, it
can be shown that
\begin{multline}\label{eq:beta1234}
\beta_{\vect q/2+\vect k\up,\vect q/2-\vect k\down,-\vect q/2+\vect k'\up,-\vect q/2-\vect k'\down} \\
=N_d\phi^{(4)}_{\vect q/2+\vect k\up,\vect q/2-\vect k\down,-\vect q/2+\vect k'\up,-\vect q/2-\vect k'\down}
\end{multline}
plus higher order corrections. Here $N_d=N/2$, and
$\phi^{(4)}$ is the momentum representation of the four-fermion wave function, describing the low-energy
scattering of two dimers. 

In coordinate space, such a four-fermion wave function
takes the asymptotic form $\propto(1-a_{dd}/R_{dd})\phi(\vect r)\phi(\vect r')$, when the distance
between dimers $R_{dd}\gg r_d$. 
Translating this long-distance boundary condition into momentum space, we get:
\begin{multline}\label{eq:phi4}
\phi^{(4)}_{\vect q/2+\vect k\up,\vect q/2-\vect k\down,-\vect q/2+\vect k'\up,-\vect q/2-\vect k'\down} \\
\approx \biggl(\delta_{\vect q,0}-\frac{4\pi a_{dd}}{\Omega q^2}\biggr)
\phi_{\vect k}\phi_{\vect k'}, ~~~q\ll 1/r_d.
\end{multline}
where $\lvert\vect k- \vect k'\rvert\gg q$, and $\Omega$ is the volume. 
%

Note that Eq.~\eqref{eq:beta_smallq} is valid in the phase space region in which $q\ll 1/r_d, \lvert \vect k-\vect k'\rvert$, while
Eq.~\eqref{eq:beta1234} is valid in the region $q, \lvert\vect k-\vect k'\rvert\gg \sqrt{n_da_{dd}}$.
Now consider the region of overlap, in which $1/r_d\gg q\gg\sqrt{n_da_{dd}}$ and $\lvert\vect k-\vect k'\rvert\gg q$.
In this region, \emph{both} the composite-boson formula [Eq.~\eqref{eq:beta_smallq}]
and the 4-fermion formula [Eq.~\eqref{eq:beta1234}]
yield exactly the \emph{same}
result, Eq.~(\ref{eq:6}) above.
Thus the conventional picture of many bosons (whose condensate suffers from a quantum depletion,
as shown by the function $x_\vect q$) 
and the nonperturbative four-fermion effects are merged into a single picture.

Finally, we turn to experimental implications of this approach
closer to unitarity.
Here we introduce higher order terms in the equation of state
[Eq.~\eqref{eq:E}] in a phenomenological manner.
Our ansatz, for which we have some
microscopic support ,
builds on earlier exact results 
\cite{Wu1959PR, Hugenholtz1959PR}
for non-composite, or true bosons. Thus
we presume that 
\begin{multline}\label{eq:E_higher}
E/\Omega=n_dE_d^{(0)}+2\pi\hbar^2 n_d^2 a_{dd}/m_d\\
\times\bigl(1+4.8144\sqrt{g}+19.6539g\ln g+C g\bigr)
\end{multline}
plus higher order terms which we shall omit below. Here
$g\equiv n_da_{dd}^3$ is the Bose gas parameter. 
The coefficient in front of $g\ln g$
was first found by Wu for the hard sphere Bose gas \cite{Wu1959PR},
and was later shown to be valid for all gases of identical non-composite bosons.

The phenomenological term containing the unknown parameter $C$ is roughly of the same order of magnitude
as the Wu term. Physically, the $C$-term must involve 
a three-dimer scattering process. This corresponds to 
a quantum \textit{six}-body problem, making $C$ very difficult to determine 
directly from theory.
For a dilute gas of  \emph{bosonic atoms} with large positive scattering length, 
a quantum \textit{three}-body calculation \cite{Braaten2002PRL}
yields $C\approx141$.
However, this latter result is not applicable to the
BEC-BCS crossover system we study here.

To proceed, we calculate $C$ from currently available experimental data,
on the radial breathing mode frequencies \cite{Kinast2004PRA}
and axial cloud sizes \cite{Bartenstein2004cloudsize}
of cigar-shaped two-component Fermi gases of $^6$Li atoms in harmonic traps.
The atomic scattering length $a$
is taken from Ref.~\cite{Bartenstein_calibration} where it is found to
diverge at $B$=834G. We presume
$a_{dd}=0.6a$ \cite{Petrov2004PRL}.
We start from Eq.~\eqref{eq:E_higher}, and evaluate the axial size 
(using the local density approximation for the density profile)
and the radial breathing mode frequency
(using the scaling ansatz of Ref.~\cite{Hu2004PRL}).
The parameter $C$ is then adjusted to best reflect the \textit{general
features} of the data.
In this way we arrive at a phenomenological equation of state
which extrapolates from deep in 
the BEC (where it is well controlled) into the range of
$k_F a$ accessible by experiment. 

\begin{figure}
\includegraphics{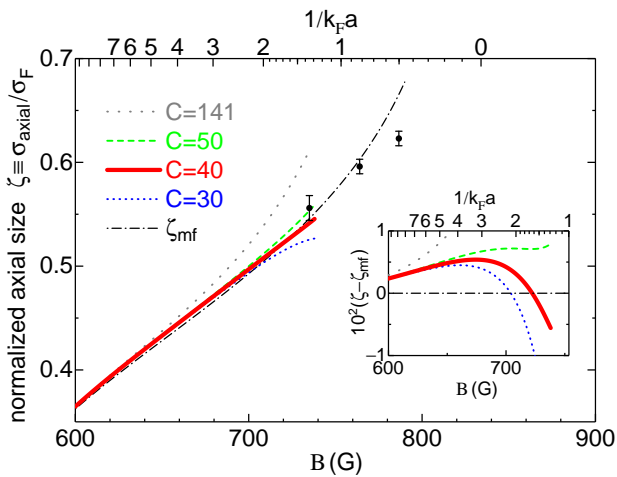}
\caption{\label{zeta_Bartenstein}
(color online). Axial cloud size $\sigma_\text{axial}$ normalized
by that of the \textit{noninteracting} Fermi gas (with the same number of
atoms), $\sigma_F$, for a cigar-shaped trapped Fermi gas of $^6$Li atoms.
Error bars: data in \cite{Bartenstein2004cloudsize}. Solid line: Eq.~\eqref{eq:E_higher} with
$C=40$. Dashed line: $C=50$. Dotted line: $C=30$. Sparse dotted line: $C=141$,
as in an \textit{atomic} BEC with large scattering length \cite{Braaten2002PRL}.
Dot-dashed line: \textit{bosonic} mean-field approximation, as in \cite{Bartenstein2004cloudsize}.
In the inset are plotted the differences between the various cloud size curves
and the bosonic mean-field curve.
}
\end{figure}

\begin{figure}
\includegraphics{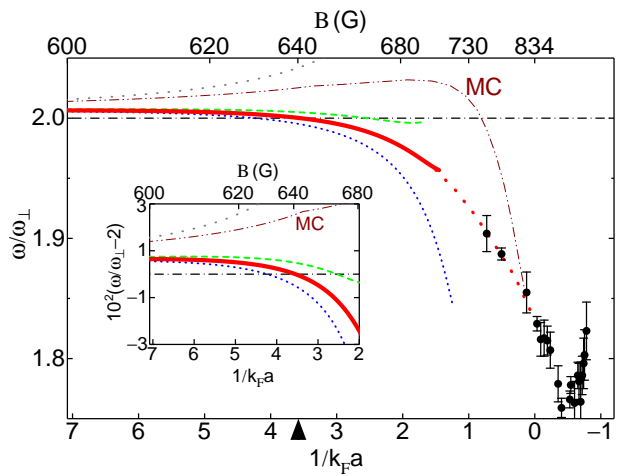}
\caption{\label{radial_Kinast}
(color online). Radial breathing mode frequency (normalized by the radial trap frequency)
versus $1/k_Fa$.
Error bars: data from \cite{Kinast2004PRA}. Solid line: Eq.~\eqref{eq:E_higher} with
$C=40$. Dashed line: $C=50$. Dotted line: $C=30$. Sparse dotted line: $C=141$,
as in an \textit{atomic} BEC with large scattering length \cite{Braaten2002PRL}.
Dot-dashed line: \textit{bosonic} mean-field approximation.
Dot-dot-dashed line: Monte-Carlo simulations \cite{Astrakharchik2005}.
The solid line is extended to indicate the likely continuous curve consistent with data, however
there is no analytic formula for this extension.
``$\blacktriangle$'' shows the $1/k_Fa$ value at which the frequency shift changes
sign, for $C=40$. In the inset are plotted the
differences between various curves and the bosonic mean-field value.}
\end{figure}

In Fig.~\ref{zeta_Bartenstein}, we plot the normalized cloud size $\zeta$ as a function
of $B$. In Fig.~\ref{radial_Kinast}, we plot the radial breathing mode frequency as
a function of $1/k_Fa$. In both figures, $k_F$ is the Fermi wave vector 
of the noninteracting Fermi gas at the trap center.
A value $C=40$ reproduces the \textit{trends} of the first generation radial mode frequencies \cite{Kinast2004PRA}
 and is also not inconsistent with the cloud size data \cite{Bartenstein2004cloudsize}.
From the cloud size data, we observe that
there may be some uncertainty in $C$ of the order $\pm 10$, so 
in both Fig.~\ref{zeta_Bartenstein} and Fig.~\ref{radial_Kinast}
we plot
$C=30$ and $C=50$ for comparison. This value of
$C$ should be compared with an
\textit{atomic} BEC with large scattering length,
for which $C\approx141$ \cite{Braaten2002PRL}.
%
%

Monte-Carlo (MC)
simulations \cite{Astrakharchik2004PRL, Astrakharchik2005}
yield a curve for the radial breathing-mode frequencies significantly higher than the published experimental data
\cite{Kinast2004PRA}.
However, there are currently ``word of mouth" indications that a
new generation of breathing mode experiments from the Innsbruck group
is in good agreement with 
MC data.
If we use the MC results to deduce $C$, we obtain
$90<C<120$ (where our result is most highly constrained 
by the MC point \cite{Astrakharchik2004PRL} at $1/k_Fa=1$). By contrast, the range of acceptable
values from the cloud size experiment is $40<C<65$
(where our result is most highly constrained by 
the data point at $B=735$G in Fig.~\ref{zeta_Bartenstein}).
That there is no overlap might be of some concern.  However, it can be shown
that the MC data itself \cite{Astrakharchik2004PRL} is not entirely consistent
(within error bars) with the cloud size data \cite{Bartenstein2004cloudsize}.
This is not unexpected since these Monte Carlo simulations, which are
based on the fixed-node approximation \cite{Astrakharchik2004PRL} 
establish an upper bound
to the equation of state;
at the same time, it is quite plausible that these first generation
cloud size experiments may be revised in due time.

It should be stressed, finally, that
the range
of validity of our \textit{low-density} formula is, strictly speaking,
outside the range
of fields accessed in the data which are closer to
unitarity. Future
experiments which
penetrate more deeply into the low density regime will be needed
to determine the parameter C
in more detail.
As a parenthetical observation, one
can see by comparing both figures that
the curves associated with different values of C in the cloud size 
experiments show a smaller deviation from each other,
than do the curves associated with the collective modes.
This is because the cloud size is proportional to the 1/5-th power of 
the dimer-dimer scattering
length \cite{Bartenstein2004cloudsize} and is thus less sensitive to the details of the equation of state.

In conclusion the
central contribution of our paper
is the proposal for  
a new many-body ground state wave function of the
atomic Fermi gas with BEC-BCS crossover. 
This wave function yields a ground state energy 
consistent with the correct dimer-dimer scattering length $a_{dd}$ in the BEC regime \cite{Petrov2004PRL}.
It also reproduces the Lee-Yang term, whose 
numerical value is also fixed by $a_{dd}$. 
We have also examined the next term beyond the Lee-Yang correction
in a phenomenological fashion, building on cloud size data
and collective mode experiments.  One has to view this phenomenological
analysis
with some caution since experiments are performed close to
unitarity whereas the theory presented here is most appropriate in
the deep BEC. 
More experiments on the cloud size and collective
mode frequencies in the deep BEC regime will be required to narrow down a
range of the parameter C
which appears in our theory,
and thus determine the appropriate equation of state. 
In the present paper we have had to make due with
experiments close to unitarity and with word of mouth information
which indicates that the measured radial breathing mode frequencies
are in a state of flux, and tending to now yield good agreement with
the Monte Carlo data.
With that proviso the value of $C$ can be as small as 40 and
as large as 120. Finally, we stress that there is
a distinction here between a BEC of fermionic dimers and
an \textit{atomic} Bose gas with large scattering length where it has been shown
\cite{Braaten2002PRL}
that in the latter $C \approx 141$.

\textbf{Note Added--}
After this work was completed we became aware of a later related study
by Levinsen and Gurarie \cite{Gurarie2006PRA}
which produced the
correct mean field term in
the equation of state, 
but which did not address the Lee-Yang term, and the other corrections
considered here.

This work is partially supported by NSF under NSF-MRSEC
Grant No. DMR-0213745.  We thank J.~E.~Thomas
and C.~Chin, as well as Q.~Chen,
E.~Braaten, A.~Bulgac, G.~Shlyapnikov and D.~S.~Petrov
for helpful communications. We thank anonymous referees of this manuscript for
some suggestions.
\bibliography{fermi24}
\end{document}